\documentclass[]{pasj01}

\Received{}
\Accepted{}
 
\begin{document} 

\title{ 
Do blue galaxy-clusters have hot intracluster gas?
}

\author{Rana \textsc{Misato}\altaffilmark{1}}
\author{Yoshiki \textsc{Toba}\altaffilmark{2,1,3,4}}
\author{Naomi \textsc{Ota}\altaffilmark{1}}
\email{naomi@cc.nara-wu.ac.jp}
\author{Naoaki \textsc{Yamamoto}\altaffilmark{5}}
\author{Tadayuki \textsc{Kodama}\altaffilmark{5}}
\author{Nobuhiro \textsc{Okabe}\altaffilmark{6,7,8}}
\author{Masamune \textsc{Oguri}\altaffilmark{9,10,11}}
\author{Ikuyuki \textsc{Mitsuishi}\altaffilmark{12}}

\altaffiltext{1}{Department of Physics, Nara Women's University, Kitauoyanishi-machi, Nara, Nara 630-8506, Japan}
\altaffiltext{2}{Department of Astronomy, Kyoto University, Kitashirakawa-Oiwake-cho, Sakyo-ku, Kyoto 606-8502, Japan}
\altaffiltext{3}{Academia Sinica Institute of Astronomy and Astrophysics, 11F of
Astronomy-Mathematics Building, AS/NTU, No.1, Section 4, Roosevelt
Road, Taipei 10617, Taiwan}
\altaffiltext{4}{Research Center for Space and Cosmic Evolution, Ehime University,
2-5 Bunkyo-cho, Matsuyama, Ehime 790-8577, Japan}
\altaffiltext{5}{Astronomical Institute, Tohoku University, 6-3, Aramaki, Aoba, Sendai, Miyagi, 980-8578, Japan}
\altaffiltext{6}{Department of Physical Science, Hiroshima University, 1-3-1 Kagamiyama, Higashi-Hiroshima, Hiroshima 739-8526, Japan}
\altaffiltext{7}{Hiroshima Astrophysical Science Center, Hiroshima University, 1-3-1, Kagamiyama, Higashi-Hiroshima, Hiroshima 739-8526, Japan}
\altaffiltext{8}{Core Research for Energetic Universe, Hiroshima University, 1-3-1, Kagamiyama, Higashi-Hiroshima, Hiroshima 739-8526, Japan}
\altaffiltext{9}{Research Center for the Early Universe, University of Tokyo, Tokyo 113-0033, Japan}
\altaffiltext{10}{Department of Physics, University of Tokyo, Tokyo 113-0033, Japan}
\altaffiltext{11}{Kavli Institute for the Physics and Mathematics of the Universe (Kavli IPMU, WPI), University of Tokyo, Chiba 277-8582, Japan}
\altaffiltext{12}{Department of Physics, Nagoya University, Aichi 464-8602, Japan}


\maketitle

\begin{abstract}
We present herein a systematic X-ray analysis of blue galaxy-clusters at $z=0.84$ discovered by the Subaru telescope. The sample consisted of 43 clusters identified by combining red-sequence and blue-cloud surveys, covering a wide range of emitter fractions (i.e., 0.3--0.8). The spatial extent of the over-density region of emitter galaxies was approximately 1~Mpc in radius. The average cluster mass was estimated as $0.6 (<1.5)\times10^{14}~{\rm M_\odot}$ from the stacked weak-lensing measurement. We analyzed the XMM-Newton archival data, and measured the X-ray luminosity of the hot intracluster medium. As a result, diffuse X-ray emission was marginally detected in 14 clusters, yielding an average luminosity of $5\times 10^{42}~{\rm erg\,s^{-1}}$. On the contrary, it was not significant in 29 clusters. The blue clusters were significantly fainter than the red-dominated clusters, and the X-ray luminosity did not show any meaningful correlation either with emitter fraction or richness. 
The X-ray surface brightness was low, but the amount of gas mass was estimated to be comparable to that observed in the $10^{13-14}~{\rm M_{\odot}}$ cluster. Based on the results, we suggest that the blue clusters are at the early formation stage, and the gas is yet to be compressed and heated up to produce appreciable X-rays. Follow-up spectroscopic measurements are essential to clarify the dynamical status and co-evolution of galaxies and hot gas in the blue clusters.
\end{abstract}


\section{Introduction}
Galaxy clusters offer the excellent laboratories for exploring how hot baryons evolve and interplay with galaxies over 10 billion years. Optical observations have shown that distant galaxy clusters have an excess of blue galaxies compared to nearby clusters. This effect is known as the Butcher-Oemler effect \citep{Butcher84}. Later works have reported that it holds up to $z\sim1$, and that the  fraction of blue galaxies depends on the cluster richness \citep{Pintos-Castro19}. In the X-ray regime, there is a longstanding question of why the X-ray luminosity of diffuse hot gas in nearby galaxy groups is anti-correlated with the spiral fraction \citep{Mulchaey03, Ota04a}. X-ray-bright groups tend to host at least one elliptical galaxy, while X-ray emission is preferentially suppressed in spiral-dominated groups. These results show a close link between galaxy morphology and cluster evolution; however, little is known about gas properties and heating mechanisms in blue clusters due to the difficulty of constructing a sample covering wide ranges of blue fraction and richness. Therefore, a large-scale cluster survey is needed to study the co-evolution of galaxies and the intracluster medium (ICM).

The Hyper Suprime-Cam \citep{Miyazaki18} Subaru Strategic Program is a wide-field imaging survey (HSC-SSP; \cite{Aihara18a,Aihara18b,Tanaka18,Bosch18}). Its area and depth provide an ideal data set for studying clusters in the distant universe. The red-sequence survey is often used to identify clusters; however, it is biased to rich clusters because it is sensitive only to passively-evolving galaxies. By combining with the blue-cloud survey (i.e., a survey of emission-line galaxies based on narrow-band filters), one can find younger, actively star-forming clusters. Yamamoto et al. (in prep.) conducted the Hybrid Search for Clusters with HSC (the HSC$^2$ survey) and found hundreds of new clusters at $0.4<z<1.7$ in the Ultra-Deep and Deep fields with a variety of emitter fractions. 

To examine whether the distant blue clusters host hot ICM and constrain the evolutionary stage, we performed a systematic X-ray analysis of clusters at $z=0.84$ discovered by the HSC$^2$ survey. The remainder of this paper is structured as follows: Sections~\ref{sec:sample} and \ref{sec:optical} represent the sample and optical data analyses of the galaxy distribution and weak-lensing mass, respectively; Section~\ref{sec:analysis} describes the XMM-Newton archival data analysis and the X-ray luminosity measurement; Section~\ref{sec:results} derives the relations between X-ray luminosity and optical observable; and Section~\ref{sec:discussion} discusses the implication of the results.

The cosmological parameters used throughout this paper are $\Omega_{m0}=0.28$, $\Omega_\Lambda=0.72$ and $h=0.7$. Accordingly, $1\arcmin$ corresponds to 463~kpc at $z=0.84$. We use the abundance table from \citet{Lodders09} in X-ray spectral modeling (section~\ref{subsec:luminosity}). The quoted errors represent the $1\sigma$ statistical uncertainties, unless stated otherwise.

\section{Sample}\label{sec:sample}
The HSC$^2$ catalog was constructed based on two kinds of surveys, that is the red-sequence and blue-cloud surveys in the Subaru HSC-SSP fields (Yamamoto et al. in prep.). The idea is to mitigate the sampling bias of the galaxy clusters to be selected, considering that the conventional red-sequence survey alone would be biased toward older clusters, in which passively evolving massive elliptical galaxies are already developed and young clusters dominated by star-forming galaxies would be missed. In the red-sequence survey, red galaxies at $z=0.8-0.9$ were selected according to the color evolution model of elliptical galaxies constructed with the stellar population synthesis model \citep{Kodama97, Kodama98} together with the photometric redshift \citep{Aihara18a, Aihara18b}. Galaxies that can be affected by bad pixels or have poor photometric measurements were removed to obtain a cleaner sample. The blue-cloud survey utilized the catalog of emission-line galaxies, specifically the [OIII] galaxies at $z=0.84$ identified by NB921 narrow-band imaging \citep{Hayashi18}. In the two surveys, the over-density regions of galaxies were then searched for by following the fixed-aperture method \citep{Toshikawa16} assuming the aperture radius of $2\arcmin.3$. In the blue-cloud survey, 55 cluster candidates at $z=0.84$ were newly found at $>3.4\sigma$ significance in the over-density of the [OIII] emitters. The full description of the HSC$^2$ catalog will be presented in a forthcoming paper (Yamamoto et al. in prep.).

\begin{figure}[htb]
\begin{center}
\includegraphics[width=8cm]{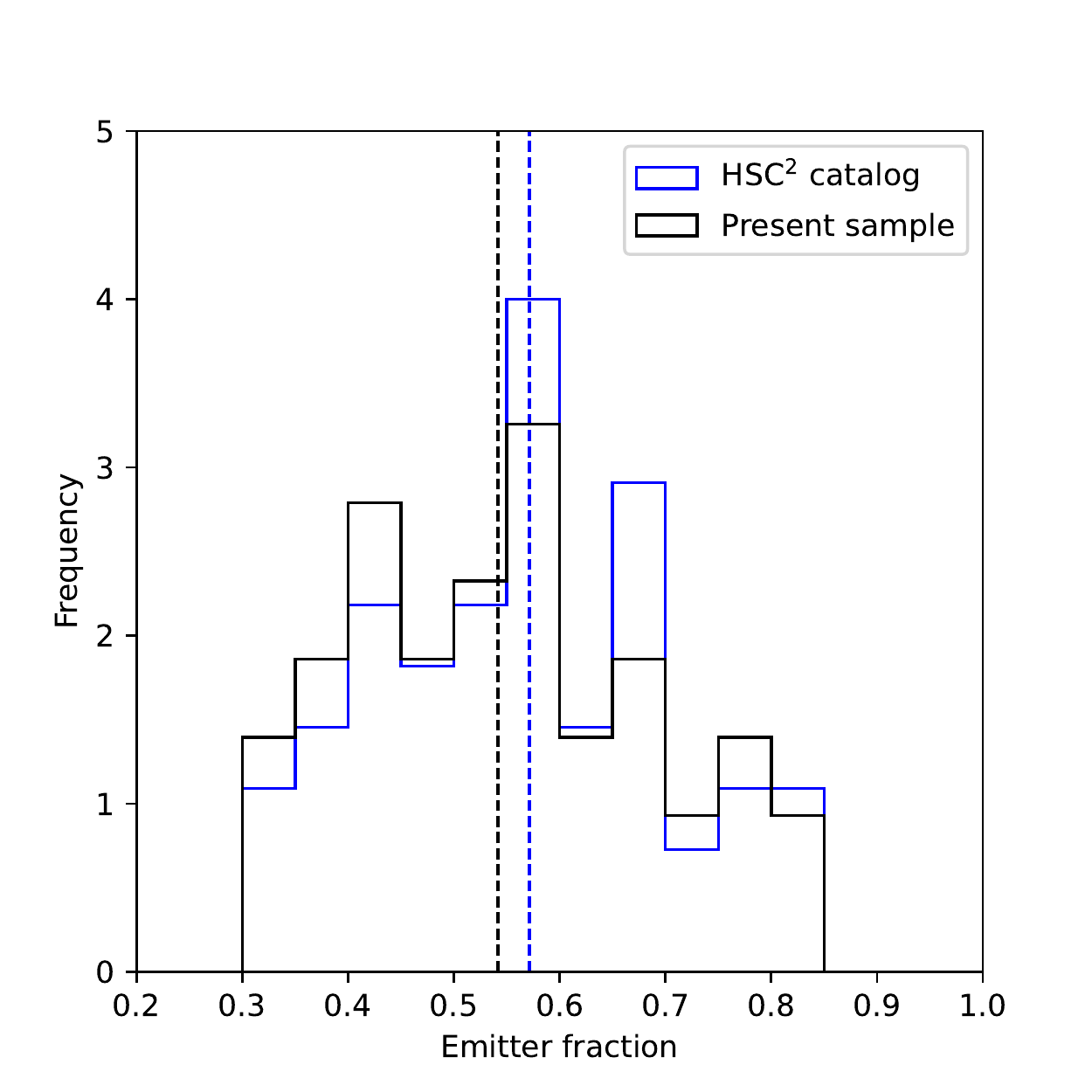}
\end{center}
\caption{Emitter-fraction distribution of the present sample (black) and the HSC$^2$ catalog (blue). The bin size is 0.05. Each histogram is normalized such that the integral over the range is unity. The vertical dashed lines indicate the medians of the present sample (black) and the HSC$^2$ catalog (blue) ($\tilde{f}_e=0.54$ and 0.57, respectively).} \label{fig:hist}
\end{figure}

\begin{figure}[htb]
\begin{center}
\includegraphics[width=8cm]{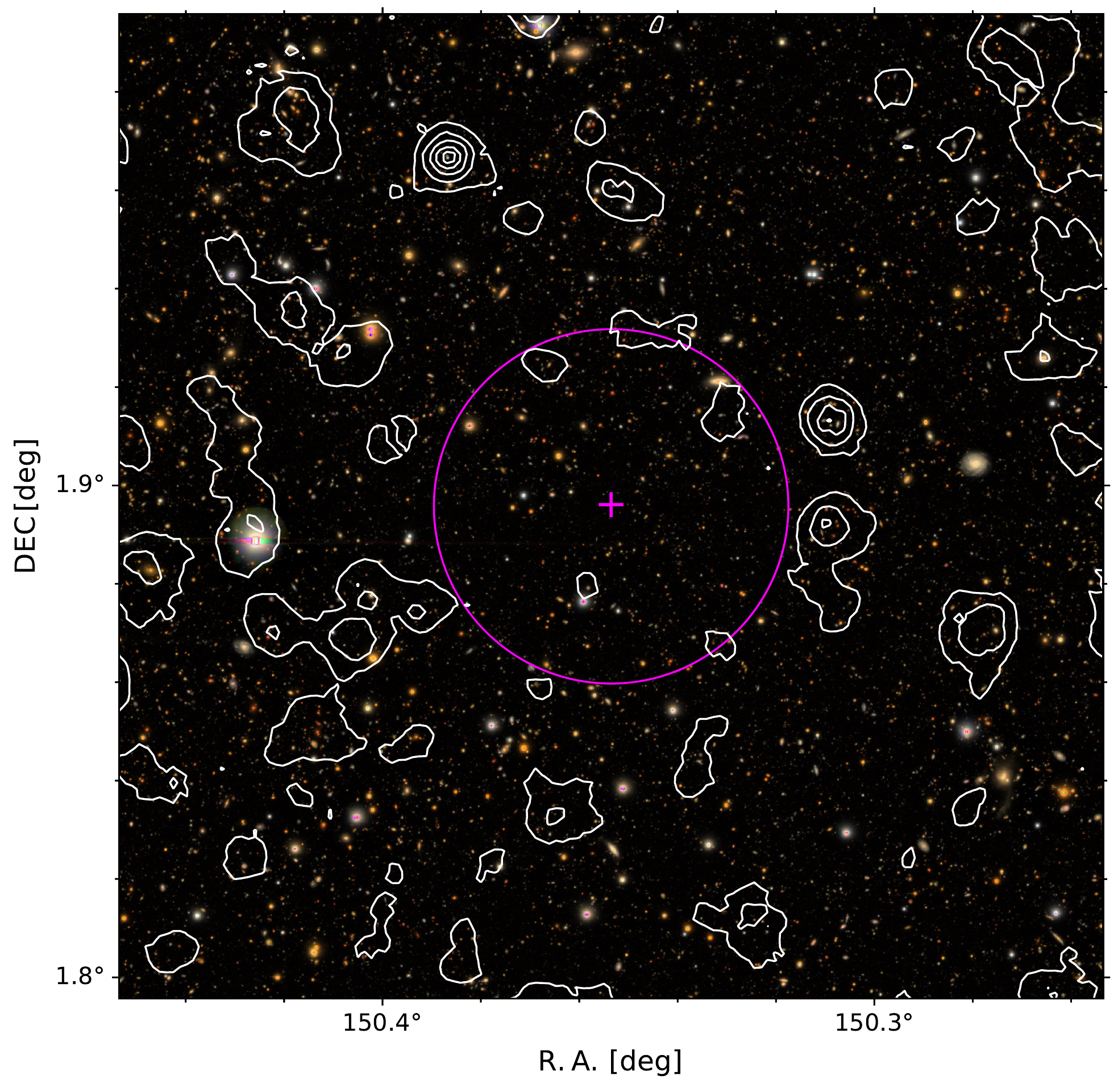}
\end{center}
\caption{Example of the HSC riz-band composite image of the blue cluster at $z=0.84$, HSC~J1001+0153. The panel size is $6\arcmin\times6\arcmin$. The white contours for the X-ray surface brightness are 15 levels logarithmically spaced from [10, 1000]~${\rm cts\,s^{-1}\,deg^{-2}}$. The optical cluster center is marked with '+'. The X-ray spectral integration radius (1~Mpc) is indicated by the magenta circle. } \label{fig:image}
\end{figure}

The HSC$^2$ catalog comprises 55 clusters at $z=0.84$. We searched the XMM-Newton Master Catalog within $1\arcmin$ around the optical centers to find that 43 clusters have the XMM-Newton archive data. Table~\ref{tab:sample} lists our sample. The richness, $N$, is defined as the total number of member galaxies, including both red and emitter galaxies identified by the red-sequence and blue cloud surveys, respectively. We define the emitter fraction, $f_e$, as the ratio of the number of emitter galaxies to $N$, whose distribution is shown in figure~\ref{fig:hist}.  $\delta_e$ represents the significance of emitter-galaxy overdensity. 
Figure~\ref{fig:image} illustrates an example of the HSC image of the blue cluster with overlaid X-ray intensity contours. 

\begin{longtable}{lp{8mm}p{8mm}p{8mm}lp{8mm}p{8mm}lll}
\caption{Sample list}\label{tab:sample}
\hline\hline
Cluster & $N^{\mathrm{a}}$ & $f_e^{\mathrm{b}}$ & $\delta_e^{\mathrm{c}}$ & Exposure$^{\mathrm{d}}$  & $N_{\rm H}^{\mathrm{e}}$ & $kT^{\mathrm{f}}$ & $L_X^{\mathrm{g}}$ & $n_e^{\mathrm{h}}$ & $M_{\rm gas}^{\mathrm{h}}$\\
\hline
\endfirsthead
\hline
Cluster & $N^{\mathrm{a}}$ & $f_e^{\mathrm{b}}$ & $\delta_e^{\mathrm{c}}$ & Exposure$^{\mathrm{d}}$  & $N_{\rm H}^{\mathrm{e}}$ & $kT^{\mathrm{f}}$ & $L_X^{\mathrm{g}}$ & $n_e^{\mathrm{h}}$ & $M_{\rm gas}^{\mathrm{h}}$\\
\hline
\endhead
\hline
\endfoot
\hline
\multicolumn{10}{l}{$^{\mathrm{a}}$ Richness.} \\
\multicolumn{10}{l}{$^{\mathrm{b}}$ Emitter fraction.}\\
\multicolumn{10}{l}{$^{\mathrm{c}}$ Overdensity of the emitter galaxies ($\sigma$).}\\
\multicolumn{10}{l}{$^{\mathrm{d}}$ Total exposure times of the three EPIC cameras (MOS1, MOS2, PN) after data screening (ks).}\\
\multicolumn{10}{l}{$^{\mathrm{e}}$ Hydrogen column density ($10^{20}~{\rm cm^{-2}}$).} \\
\multicolumn{10}{l}{$^\mathrm{f}$ ICM temperature (keV) estimated from the richness $N$ and the $N-T$ relation.}\\
\multicolumn{10}{l}{$^\mathrm{g}$ Rest-frame 0.5--2~keV luminosity within 1~Mpc ($10^{43}{\rm erg\,s^{-1}}$).}\\
\multicolumn{10}{l}{$^\mathrm{h}$ Average electron density ($10^{-4}~{\rm cm^{-3}}$) and gas mass within 1~Mpc ($10^{12}{\rm M_{\odot}}$).}\\
\endlastfoot
HSC~J1002+0248 & 24 & 0.71 & 4.4 &  29.2, 43.1, 30.3 	& 1.80 & 2.7 & $ 0.96 \pm 0.83 $  & $ 2.3 \pm 0.4 $ & $ 3.3 \pm 0.6$ \\ 
HSC~J2328-0055 & 21 & 0.57 & 3.4 &   0.0,  3.8,  0.0 	& 4.55 & 2.5 & $ < 7.84 $ & $< 0.7$  & $< 0.9$ \\ 
HSC~J2323-0048 & 19 & 0.84 & 4.9 &  25.9, 27.2, 18.9 	& 3.66 & 2.4 & $ < 1.64 $ & $< 0.2$  & $< 0.2$ \\ 
HSC~J2327-0045 & 22 & 0.68 & 3.9 &   0.0,  5.6,  0.0 	& 4.42 & 2.6 & $ 12.24 \pm 5.63 $  & $ 4.5 \pm 1.0 $ & $ 6.1 \pm 1.4$ \\ 
HSC~J2326-0031 & 17 & 0.71 & 3.4 &   1.6,  4.1,  0.0 	& 4.18 & 2.3 & $ 16.40 \pm 11.30 $  & $ 7.2 \pm 1.8 $ & $ 7.8 \pm 2.0$ \\ 
HSC~J2329-0027 & 21 & 0.76 & 4.9 &   3.9,  4.0,  0.0 	& 4.19 & 2.5 & $ < 4.71 $ & $< 0.4$  & $< 0.6$ \\ 
HSC~J2327-0025 & 28 & 0.50 & 4.9 &   6.2,  9.2,  2.5 	& 4.12 & 2.9 & $ 1.81 \pm 2.76 $  & $ 3.0 \pm 0.9 $ & $ 5.0 \pm 1.5$ \\ 
HSC~J2327-0022 & 20 & 0.60 & 3.9 &   4.5,  5.1,  2.5 	& 4.05 & 2.5 & $ 6.79 \pm 2.41 $  & $ 3.9 \pm 0.6 $ & $ 4.9 \pm 0.7$ \\ 
HSC~J2327-0018 & 22 & 0.64 & 4.9 &   6.1,  9.2,  0.8 	& 4.00 & 2.6 & $ 1.33 \pm 3.22 $  & $ 4.3 \pm 1.0 $ & $ 5.9 \pm 1.4$ \\ 
HSC~J2328-0018 & 19 & 0.68 & 3.4 &   7.9,  9.2,  4.8 	& 3.91 & 2.4 & $ < 1.90 $ & $< 0.2$  & $< 0.3$ \\ 
HSC~J2329-0010 & 41 & 0.32 & 4.4 &   6.8, 11.8,  4.7 	& 4.05 & 3.5 & $ < 1.98 $ & $< 0.2$  & $< 0.5$ \\ 
HSC~J2328-0006 & 20 & 0.60 & 3.9 &   3.4,  8.8,  2.1 	& 3.90 & 2.5 & $ < 2.07 $ & $< 0.3$  & $< 0.3$ \\ 
HSC~J2328+0002 & 15 & 0.80 & 3.9 &   0.0, 35.9,  0.0 	& 4.21 & 2.1 & $ < 1.64 $ & $< 0.2$  & $< 0.2$ \\ 
HSC~J2329+0009 & 20 & 0.75 & 4.9 &   0.0, 35.9,  0.0 	& 4.21 & 2.5 & $ < 1.09 $ & $< 0.2$  & $< 0.2$ \\ 
HSC~J0959+0138 & 28 & 0.61 & 4.9 &  64.2, 79.3, 55.4 	& 1.79 & 2.9 & $ < 0.54 $ & $< 0.1$  & $< 0.1$ \\ 
HSC~J1002+0153 & 29 & 0.59 & 4.9 &  53.6, 81.4, 58.8 	& 2.04 & 3.0 & $ < 0.54 $ & $< 0.1$  & $< 0.2$ \\ 
HSC~J1001+0153 & 21 & 0.52 & 3.4 & 220.6,224.1,125.1 	& 1.83 & 2.5 & $ < 0.41 $ & $< 0.1$  & $< 0.1$ \\ 
HSC~J1001+0153 & 38 & 0.47 & 5.9 &  72.8,109.9, 61.5 	& 1.87 & 3.4 & $ < 0.41 $ & $< 0.1$  & $< 0.1$ \\ 
HSC~J1001+0159 & 39 & 0.36 & 4.9 &  56.0, 56.7, 40.1 	& 1.85 & 3.4 & $ < 0.49 $ & $< 0.1$  & $< 0.2$ \\ 
HSC~J1001+0202 & 37 & 0.41 & 4.4 & 162.7,166.5, 84.6 	& 1.78 & 3.4 & $ 0.14 \pm 0.45 $  & $ 4.2 \pm 0.3 $ & $ 8.9 \pm 0.6$ \\ 
HSC~J1001+0203 & 41 & 0.39 & 5.4 & 208.3,240.8,149.9 	& 1.76 & 3.5 & $ < 0.28 $ & $< 0.0$  & $< 0.1$ \\ 
HSC~J1001+0207 & 48 & 0.54 & 7.4 & 125.7,139.3, 76.7 	& 1.74 & 3.8 & $ < 0.40 $ & $< 0.1$  & $< 0.2$ \\ 
HSC~J1001+0208 & 62 & 0.50 & 10.9 & 137.2,158.9, 93.3 	& 1.75 & 4.3 & $ < 0.32 $ & $< 0.1$  & $< 0.2$ \\ 
HSC~J1002+0209 & 44 & 0.39 & 4.9 & 104.4,107.0, 38.4 	& 1.78 & 3.7 & $ < 0.47 $ & $< 0.1$  & $< 0.2$ \\ 
HSC~J1000+0212 & 46 & 0.46 & 6.9 & 148.0,153.5, 94.3 	& 1.72 & 3.7 & $ 0.68 \pm 0.36 $  & $ 2.2 \pm 0.2 $ & $ 5.6 \pm 0.5$ \\ 
HSC~J0957+0215 & 39 & 0.49 & 6.9 &  40.3, 40.1, 21.9 	& 1.81 & 3.4 & $ < 0.47 $ & $< 0.1$  & $< 0.2$ \\ 
HSC~J0958+0214 & 41 & 0.32 & 4.4 &  69.1, 77.1, 37.4 	& 1.76 & 3.5 & $ 0.64 \pm 0.39 $  & $ 2.2 \pm 0.3 $ & $ 5.0 \pm 0.6$ \\ 
HSC~J1002+0213 & 26 & 0.46 & 3.4 &  96.2,113.3, 36.0 	& 1.80 & 2.8 & $ < 0.31 $ & $< 0.1$  & $< 0.1$ \\ 
HSC~J0958+0216 & 27 & 0.56 & 3.9 & 139.3,139.3, 91.4 	& 1.73 & 2.9 & $ 0.08 \pm 0.39 $  & $ 2.0 \pm 0.3 $ & $ 3.3 \pm 0.5$ \\ 
HSC~J1002+0219 & 19 & 0.58 & 3.9 & 130.0,182.5, 72.7 	& 1.80 & 2.4 & $ < 0.35 $ & $< 0.1$  & $< 0.1$ \\ 
HSC~J0957+0226 & 34 & 0.38 & 3.9 &  36.8, 37.1, 16.9 	& 1.85 & 3.2 & $ < 0.91 $ & $< 0.1$  & $< 0.2$ \\ 
HSC~J1002+0240 & 19 & 0.63 & 4.4 &  70.6, 72.7, 49.6 	& 1.78 & 2.4 & $ < 0.38 $ & $< 0.1$  & $< 0.1$ \\ 
HSC~J1001+0250 & 33 & 0.42 & 4.9 &  39.5, 39.4, 30.3 	& 1.73 & 3.2 & $ 2.46 \pm 1.54 $  & $ 8.5 \pm 0.5 $ & $ 16.4 \pm 1.0$ \\ 
HSC~J0219-0522 & 28 & 0.43 & 3.9 &  72.1, 97.2, 60.9 	& 2.10 & 2.9 & $ < 0.43 $ & $< 0.1$  & $< 0.1$ \\ 
HSC~J0220-0458 & 18 & 0.78 & 3.4 &  62.8, 86.3, 20.9 	& 1.95 & 2.3 & $ < 0.33 $ & $< 0.1$  & $< 0.1$ \\ 
HSC~J0220-0454 & 49 & 0.35 & 5.9 &  86.1, 86.9, 39.6 	& 1.98 & 3.9 & $ 0.51 \pm 0.35 $  & $ 1.5 \pm 0.2 $ & $ 4.1 \pm 0.6$ \\ 
HSC~J0218-0447 & 36 & 0.44 & 3.9 &  10.5, 41.8, 19.4 	& 1.95 & 3.3 & $ 0.44 \pm 0.89 $  & $ 2.4 \pm 0.6 $ & $ 5.0 \pm 1.2$ \\ 
HSC~J0221-0447 & 20 & 0.65 & 4.9 &  68.3,120.5, 60.1 	& 1.86 & 2.5 & $ < 0.47 $ & $< 0.1$  & $< 0.1$ \\ 
HSC~J0218-0443 & 32 & 0.41 & 5.4 &  57.3, 67.1, 39.2 	& 1.95 & 3.1 & $ < 0.57 $ & $< 0.1$  & $< 0.2$ \\ 
HSC~J0219-0436 & 24 & 0.58 & 4.4 &  63.0, 71.6, 36.0 	& 2.03 & 2.7 & $ < 0.65 $ & $< 0.1$  & $< 0.1$ \\ 
HSC~J0219-0433 & 23 & 0.65 & 4.4 &  52.0,109.8, 58.1 	& 1.99 & 2.6 & $ < 0.50 $ & $< 0.1$  & $< 0.1$ \\ 
HSC~J0218-0429 & 36 & 0.53 & 5.9 &  44.5, 88.7, 55.5 	& 1.93 & 3.3 & $ < 0.81 $ & $< 0.1$  & $< 0.1$ \\ 
HSC~J0220-0429 & 29 & 0.45 & 4.4 &  72.1,118.3, 61.9 	& 2.01 & 3.0 & $ 0.34 \pm 0.44 $  & $ 2.2 \pm 0.3 $ & $ 3.7 \pm 0.5$ \\ 
\end{longtable}

\section{Optical data analysis}\label{sec:optical}
\subsection{Spatial distribution of emitter galaxies}\label{subsec:galaxy_dist}
To evaluate the spatial extent of the blue clusters, we calculated two kinds of average radial profiles by stacking the surface densities of the emitter galaxies and red galaxies separately. We used the same emitter galaxies defined in Section~\ref{sec:sample}. The red galaxies were selected by a photometric redshift slice with $|z-0.84|<0.05$ and $z_{\rm mag}<23$ using the MISUKI catalog \citep{Tanaka15, Tanaka18}. The emitter galaxies were excluded if they overlapped. 

\begin{figure}[htb]
\begin{center}
\includegraphics[width=8cm]{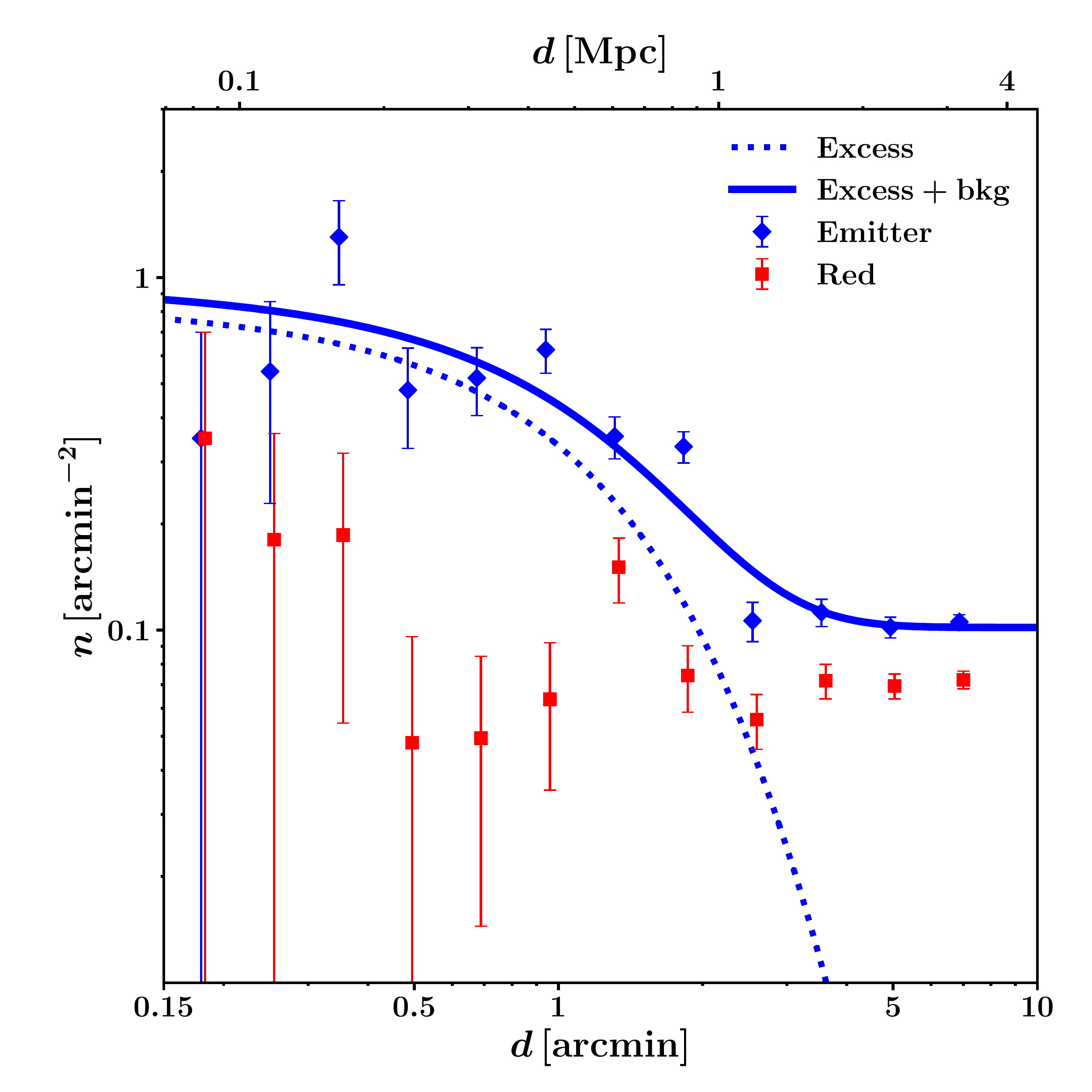}
\end{center}
\caption{Stacked surface density profiles of the emitter (blue) and red (red) galaxies in 43 blue clusters. The best-fit model and the excess component of the emitter galaxies are illustrated by the blue solid and dotted curves, respectively. } \label{fig:n_vs_r}
\end{figure}

Figure~\ref{fig:n_vs_r} shows the resulting surface density profiles. The excess of the emitter galaxies is more prominent than that of the red galaxies. The observed profile was well fit by the following mathematical form: $n(r)=n_0(1-\tanh{(r/\sigma)}) + const.$ The best-fit parameters were $n_0=0.45\pm0.11~{\rm cm^{-2}}$, $\sigma=1.20\pm0.10$~Mpc, and $const=0.102\pm0.004~{\rm cm^{-2}}$. Therefore, we estimated the average spatial extent of the emitter galaxies as 1~Mpc within which the excess was over the $3\sigma$ level.

\subsection{Weak-lensing mass measurement}\label{subsec:weaklens}
We measure average weak-lensing (WL) masses of blue clusters 
using a method of Point Spread Function (PSF) correction
known as re-Gaussianization \citep{Hirata03}, which is implemented in the HSC pipeline (see details in \cite{HSCWL1styr}) We refer to the three-year shape catalog \citep{2021arXiv210700136L}, and use only galaxies that satisfy full-color and full-depth condition from the HSC galaxy catalog to achieve both precise shape measurement and photometric redshift estimation. The number of the blue clusters of which mass can be measured is 41 out of 43. We select background galaxies behind each cluster in the color-color space, following \citet{Medezinski18}. The details of the formulation are described in \citep{2019PASJ...71...79O}. The signal-to-noise ratios for the WL signals for the whole blue clusters are $2.27$ and $2.44$ with and without the lensing covariance matrix from the uncorrelated large-scale structure, respectively. Because of the low signal-to-noise ratio, we cannot constrain the concentration parameter and thus assume the mass-concentration relation \citep{Diemer15}. The resulting mass is $M_{500}=0.11_{-0.04}^{+0.11}\times10^{14}M_\odot$. When we divide the sample into two subsamples by the threshold of emitter-richness, $N_e=15$, the mass for the lower and higher richness subsamples are $M_{500}<0.21\times10^{14}M_\odot$ and $0.15_{-0.07}^{+0.20}\times10^{14}M_\odot$, respectively.
 
\section{X-ray data analysis}\label{sec:analysis}
\subsection{Data reduction}
We retrieved the observation data files from the XMM-Newton Science Archive and reprocessed the EPIC data with the XMM-Newton Science Analysis System v17.0.0 and the Current Calibration Files. Initial data reduction, including flare screening, point source detection, and quiescent background level estimation, was performed in the standard manner by using the XMM-Newton Extended Source Analysis Software \citep{Snowden08}. We estimated the residual soft proton contamination by calculating the count-rate ratio between in-FOV and out-FOV and excluded the files, for which data were extremely contaminated (i.e., $F_{\rm in}/F_{\rm out}>1.5$) \citep{Deluca04}.

\subsection{Measurement of the X-ray luminosity}\label{subsec:luminosity}
\begin{figure}[htb]
\begin{center}
\includegraphics[width=8cm]{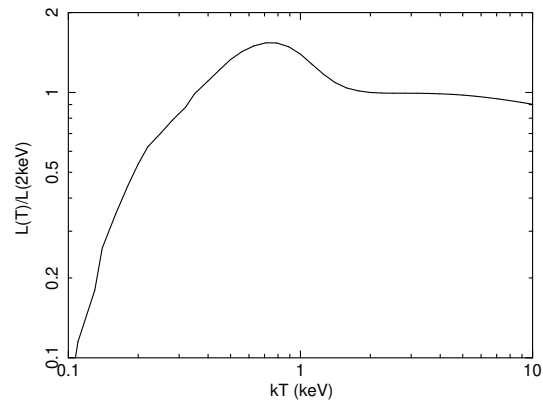}
\end{center}
\caption{Model luminosity normalized at $kT=2$~keV. The 0.5--2~keV luminosity is calculated under the assumption of the APEC model and $Z=0.3$~solar.} \label{fig:emissivity}
\end{figure}

We extracted the X-ray spectra from a circular region within $r=1$~Mpc to derive the X-ray luminosity of the hot ICM in the rest-frame 0.5--2~keV range, $L_X$. The radius was chosen based on the spatial extent of the galaxy distribution (Section~\ref{subsec:galaxy_dist}). The background was estimated from an $r=2-3$~Mpc annulus centered on the cluster center. The point sources detected by the initial data reduction and those in the 4MM-DR10 catalog \citep{Webb20} were excluded. The photon statistics of the observed spectra were low in most of the clusters; hence, we estimated $L_X$ from the EPIC count rates in the 0.5--2.0~keV band assuming the APEC thin-thermal plasma model \citep{Smith01, Foster12} and the metal abundance of 0.3~solar. We assumed the ICM temperature estimated from the richness-temperature relation of the optically-selected clusters \citep{Oguri18}. The X-ray emissivity of the APEC model weakly depends on the temperature in the soft X-ray band; therefore, the model luminosity changed only within 10\% (a factor of 2) as long as $2<kT/{\rm keV} <10$ ($0.2<kT/{\rm keV} <10$) (figure~\ref{fig:emissivity}). The Galactic absorption was corrected by the phabs model and the hydrogen column density \citep{Kalberla05}. The missing flux caused by the point-source removal was corrected by interpolating the ICM emission, assuming that the cluster surface brightness was approximately constant based on the staking analysis (see section~\ref{subsec:sb}).

We took the weighted average of the luminosities derived by the three EPIC cameras. Consequently, diffuse X-ray emission was marginally detected in 14 clusters, while no significant emission was found in 29 of the sample. For the latter, the $1\sigma$ upper limit on the cluster luminosity was derived by the statistical uncertainty of the background. Table~\ref{tab:sample} presents the results.

We evaluated the systematic error in the X-ray luminosity caused by the contamination from active galactic nuclei (AGNs). The angular size of the cluster core at $z=0.84$ can be comparable to the point-spread functions of the XMM-Newton telescopes; thus, we needed to carefully assess the AGN emission. Referring to the SDSS DR15 catalog \citep{Aguado19}, background AGNs existed within a circle of 1~Mpc in seven blue clusters. We estimated the rest-frame luminosity at 2500A to fit the SDSS spectra by using the Quasar Spectral Fitting package (QSFit; \cite{Calderone17}). Please also see \citet{Toba21}. This yielded an AGN flux that was comparable to the total X-ray flux within $R_{500}$. Therefore, the regions of these background AGNs were securely masked in the cluster-luminosity measurement.

 \subsection{Measurement of the X-ray surface brightness}\label{subsec:sb}
 We derived the average surface brightness profile to study the spatial distribution of the ICM by stacking the EPIC MOS data. We extracted the radial profiles in the 0.4--2~keV band centered on the cluster coordinates. The radial bin size was $4\arcsec$. After correcting for the exposure maps, we took their weighted average of 43 clusters. To compare it with the red-dominated clusters, we analyzed the XMM-Newton data of four clusters at $0.83<z<0.88$ taken as a part of the X-ray follow-up project of the high-redshift, massive ($N>40$) CAMIRA clusters (Ota et al. in prep.)

\begin{figure}[htb]
\begin{center}
\includegraphics[width=8cm]{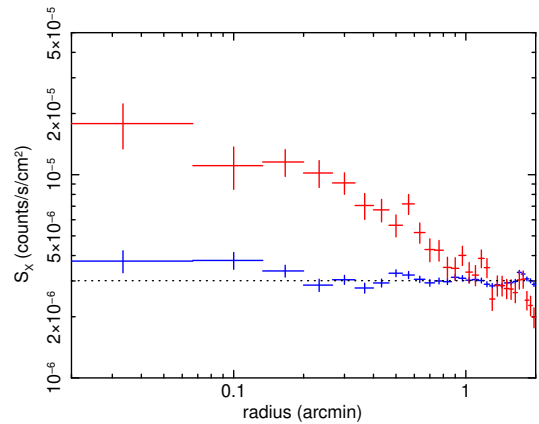}
\end{center}
\caption{Stacked surface brightness profiles of 43 blue clusters at $z=0.84$ (blue crosses) and 4 red clusters at $z\sim0.8$ (red crosses). The EPIC-MOS data in the 0.4--2~keV band are corrected for the exposure maps. The average background level estimated from the outer regions of the blue clusters ($1.3<r/{\rm arcmin}<2.0$) is indicated by the horizontal dotted line. $1\arcmin$ corresponds to 455~kpc at $z=0.8$.} \label{fig:sb}
\end{figure}

Figure~\ref{fig:sb} illustrates the resulting X-ray brightness profiles of the blue and CAMIRA clusters at comparable redshifts. The present sample showed a significantly low, and flat profile without any prominent central core, while the latter had a centrally-peaked profile as often observed in X-ray clusters (e.g., \cite{Ota04b}).

\section{Results}\label{sec:results}
 
 \subsection{Relation between X-ray luminosity and emitter fraction}
Based on the XMM-Newton data analysis in section~\ref{sec:analysis}, the blue clusters were fairly faint in the X-ray regime; No significant emission was found in 29 clusters, and the X-ray emission was only marginally detected in 14 clusters. The average luminosity of the latter 14 objects was $(0.50\pm0.15)\times10^{43}~{\rm erg\,s^{-1}}$, which was lower by a factor of 4 compared to that observed in the CAMIRA clusters \citep{Oguri18} and was rather comparable to poor groups \citep{Lovisari21} and bright elliptical galaxies \citep{Babyk18}. 

\begin{figure*}[htb]
\begin{center}
\includegraphics[width=8cm]{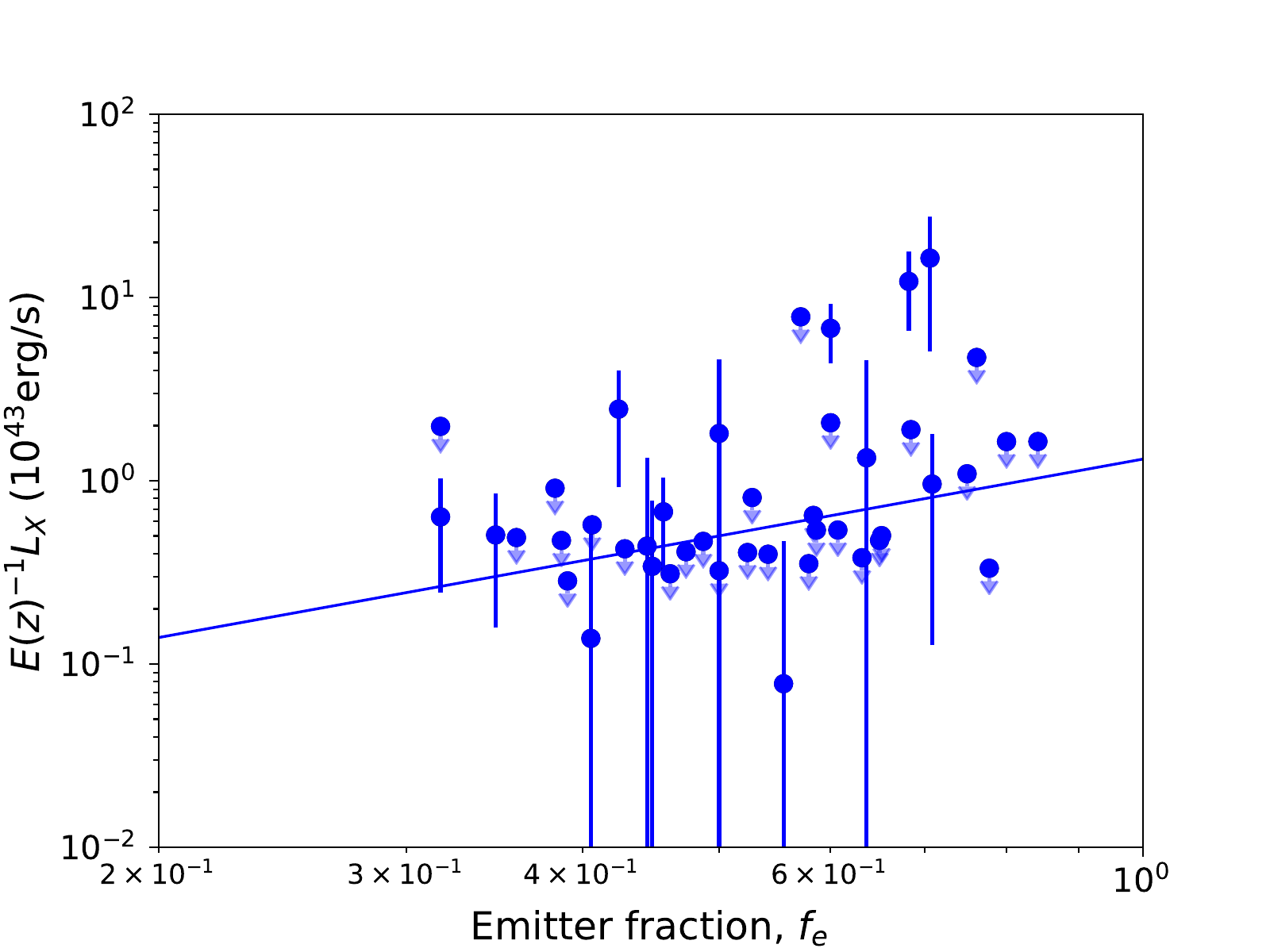}
\includegraphics[width=8cm]{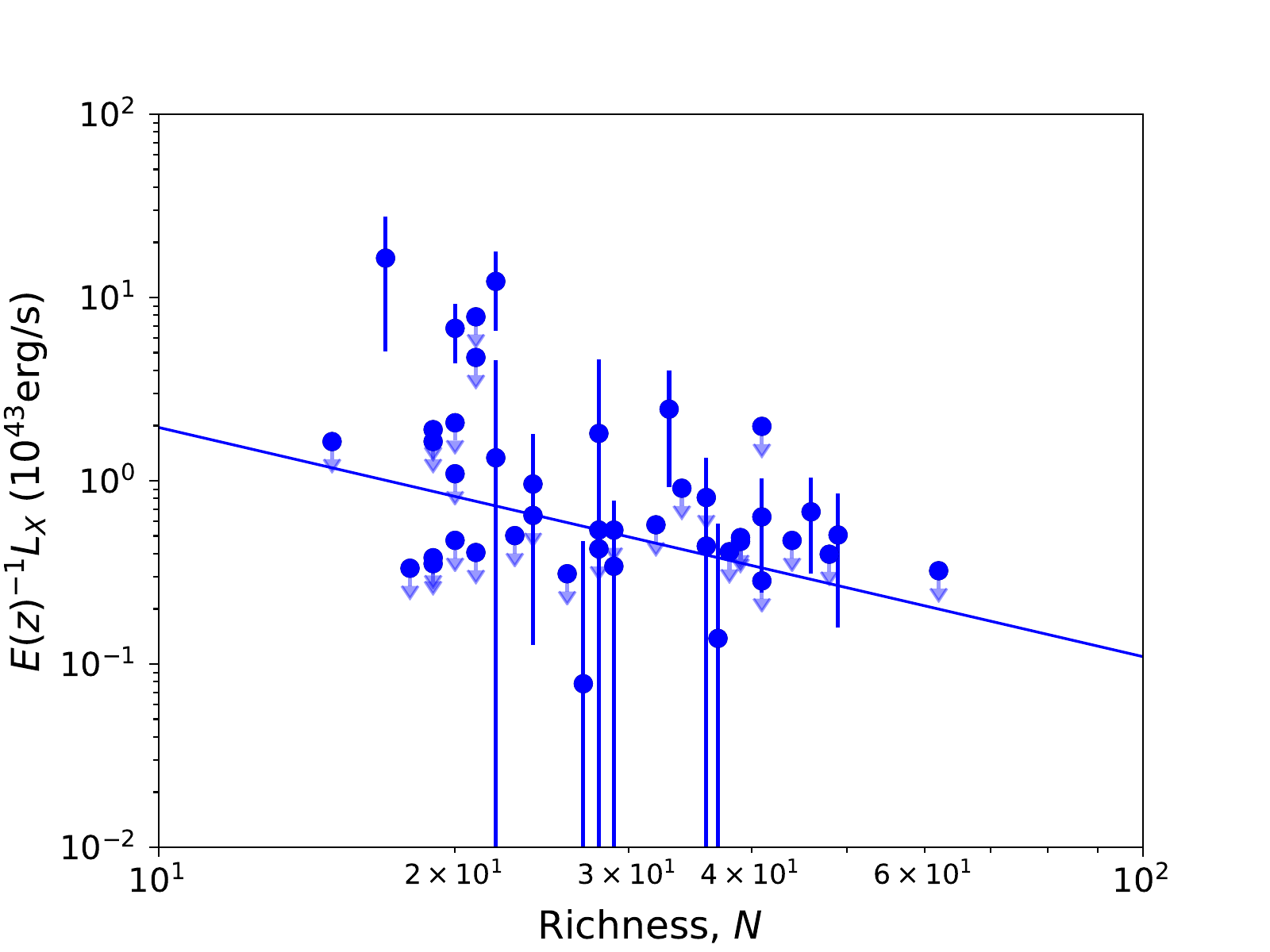}
\end{center}
\caption{(Left) Relation between the X-ray luminosity and the emitter fraction in 43 blue clusters. The circles depict the 0.5--2~keV luminosity within 1~Mpc. The error bars/arrow indicate 68\% statistical uncertainties/68\% upper limits. The solid line shows the best-fit power-law model. (Right) Relation between luminosity and richness. The meanings of symbols and the line are the same as those in the left panel.} \label{fig:lx-fb}
\end{figure*}

Figure~\ref{fig:lx-fb} shows the X-ray luminosity as a function of the emitter fraction. The evolution-corrected luminosity $E(z)^{-1}L_X$ was plotted for a comparison with the optical clusters at various redshifts \citep{Oguri18} assuming the self-similar model (e.g., \cite{Giles16}). Here $E(z) = (\Omega_M(1+z)^3 + \Omega_{\Lambda})^{1/2}$. We fit the observed $L-f_e$ relation to the power-law model (equation~\ref{eq:lx-fb}) by using the Bayesian regression method \citep{Kelly07} because it can constrain the parameters, even when the data have large measurement errors or only upper limits. The quantities $\alpha$ and $\beta$, and the intrinsic scatter $\sigma_{L|f}$ were treated as free parameters.

\begin{equation}
\log{ \left( \frac{E(z)^{-1}L_X}{10^{43}~{\rm erg\,s^{-1}}}\right) }  = \alpha  +  \beta \log{f_e}  \label{eq:lx-fb}
\end{equation}
The best-fit parameters were $\alpha=0.12\pm0.19$, $\beta=1.39\pm0.63$, and $\sigma_{L|f}=0.19\pm0.05$. The correlation coefficient was $0.35\pm0.15$, suggesting the absence of a clear correlation.

\subsection{Relation between X-ray luminosity and richness}\label{subsec:lt}
The right panel of figure~\ref{fig:lx-fb} shows the X-ray luminosity as a function of the number of richness. We fit the relation assuming the following model to quantify the correlation:

\begin{equation}
\log{\left(
 \frac{ E(z)^{-1}L_X }{ 10^{43}~{\rm erg\,s^{-1}} } \right) }  = \alpha  +  \beta \log{\left( \frac{N_r}{30}\right)}  \label{eq:lx-nr}
\end{equation}

The best-fit parameters were $\alpha=-0.31\pm0.07$, $\beta=-1.25\pm0.46$, and $\sigma_{L|N}=0.18\pm0.05$. The correlation coefficient was $-0.41\pm0.14$. Thus the present sample showed no clear $L_X-N$ correlation in contrast to the CAMIRA clusters with a positive correlation \citep{Oguri18}.

\subsection{Relation between X-ray luminosity and mass}
Next, we study the relationship between X-ray luminosity and WL mass. The stacked X-ray luminosity was computed with a weight of lensing contribution at each cluster \citep{2019PASJ...71...79O} considering that the WL masses were sensitive to the lensing efficiency and the number of background galaxies. The average $L_X$ and $M_{500}$ for the blue clusters were calculated for the whole sample as well as two sub-samples defined by the ranges of the emitter richness (the number of emitter galaxies) (i.e., 10--15 and 15--31). Similarly, the sample for the 50 CAMIRA clusters with X-ray counterparts in the XXL and XMM-LSS fields \citep{Oguri18,Clerc14,Pacaud16,Giles16} was subdivided into five by richness class (i.e., 10--19, 19--25, 25--35, 35--50, and 50--). Excluding one cluster for which the shape catalog was unavailable because it lay near the edge of the survey field \citep{2021arXiv210700136L}, we used 49 CAMIRA clusters in the analysis. The luminosities were measured in different radii (1~Mpc/$R_{500}$ for the blue clusters/the CAMIRA clusters), but we plotted their values without correction because X-ray emission was not significant outside $R_{500}\sim 1\arcmin$ in the blue clusters (figure~\ref{fig:sb}).

\begin{figure}[htb]
\begin{center}
\includegraphics[width=8cm]{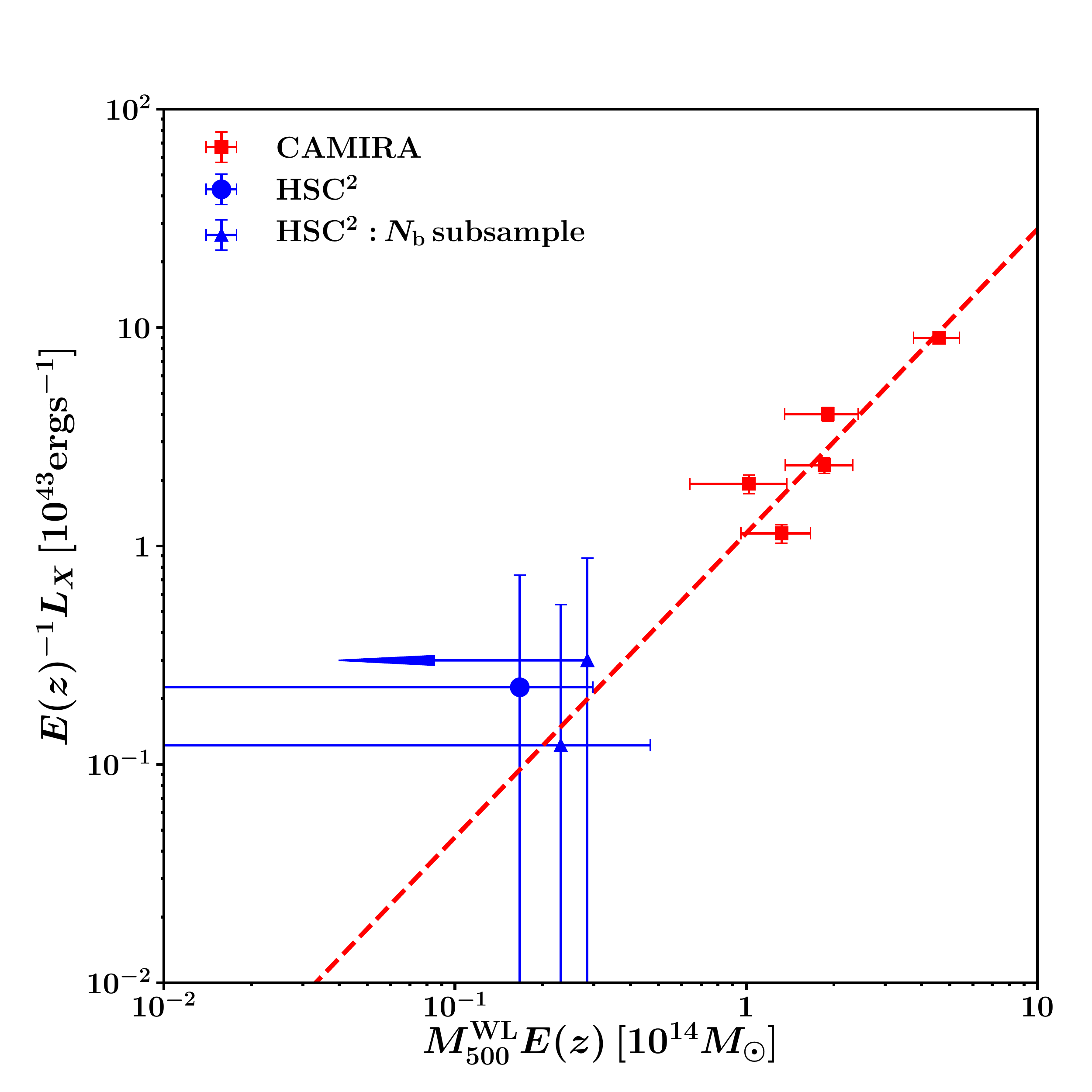}
\end{center}
\caption{X-ray luminosity-mass relation. The average 0.5--2.0~keV luminosity of the blue clusters is plotted against $M_{500}$ measured by the stacked weak-lensing analysis (blue circle). The results of the two subsamples are also plotted (blue triangles). The five subsamples of the CAMIRA clusters in the XXL and XMM-LSS catalogs are shown for comparison (red boxes). The dashed line represents the best-fit power-law model to the CAMIRA clusters.} \label{fig:lx-m}
\end{figure}

Figure~\ref{fig:lx-m} shows the $L_X-M_{500}$ relation. The CAMIRA clusters exhibited a positive correlation. The linear regression method \citep{2021arXiv211110080A} resulted in the best-fit power-law model of $E(z)^{-1}(L_X/10^{43}\,{\rm ergs^{-1}}) = 10^{a}(M_{500}E(z)/10^{14}M_\odot)^{b}$, $a=0.05_{-0.19}^{+0.17}$ and $b=1.40_{-0.44}^{+0.45}$. In contrast, the upper bounds on the average luminosity and mass were obtained for the blue clusters, and the data points of the two subsamples overlapped with each other. The present sample seemed to follow the $L_X-M$ relation of the CAMIRA clusters within the statistical errors. 

\section{Discussion}\label{sec:discussion}
The systematic analysis of the XMM-Newton data showed that diffuse X-ray emission from the blue clusters at $z=0.84$ was significantly faint compared to the optically-selected clusters in the CAMIRA catalog. We found no clear correlation between the X-ray luminosity and the emitter fraction. This trend is likely to be different from the anti-correlation seen in nearby galaxy groups \citep{Mulchaey03, Ota04a}. In what follows, we estimate the gas density and mass and discuss the evolutionary stage of the blue clusters.

\subsection{Gas properties of the blue clusters}
We estimate the electron density, $n_e$, and the gas mass, $M_{\rm gas}$ to further examine the physical status of the ICM. The stacking analysis suggested a very diffuse ICM distribution (figure~\ref{fig:sb}); thus, we simply assumed a uniform sphere of isothermal gas with a radius of 1~Mpc.
Correspondingly, the following values are meaningful only as order-of-magnitude estimates. The soft X-ray luminosity is insensitive to temperature (figure~\ref{fig:emissivity}), and the emission measure changes as $\propto n_e^2$; therefore, the temperature uncertainty will only slightly affect the present density estimate. Table~\ref{tab:sample} summarizes the results. 

For the 14 clusters with marginal X-ray detection, we obtained the average electron density as $n_e \sim 4\times10^{-4}~{\rm cm^{-3}}$. This led to the gas mass and the gas-mass fraction of $M_{\rm gas}\sim 6\times10^{12}~{\rm M_{\odot}}$ and $f_{\rm gas}=M_{\rm gas}/M_{500}\sim 0.1$ at $R_{500}= 0.53$~Mpc if we quote the WL mass of $M_{500}=0.6\times10^{14}~{\rm M_{\odot}}$ (section~\ref{subsec:weaklens}).
Therefore, we suggest that these clusters retain a gas mass comparable to that observed in a $10^{13-14}~{\rm M_{\odot}}$ cluster (e.g., \cite{Eckert21}). However, the gas distribution is yet to be concentrated to produce a strong X-ray emission.

The upper limit for the 29 undetected clusters was obtained as $n_e <10^{-5\sim-4}~{\rm cm^{-3}}$. This yielded loose upper limits on the gas mass and the gas-mass fraction of $M_{\rm gas}<10^{11\sim12}~{\rm M_{\odot}}$ and $f_{\rm gas}\lesssim 0.01$, respectively. The low density also translated into the baryon overdensity $\delta \equiv n_{H}/\bar{n_{\rm H}} \lesssim 10-100$, where $n_{\rm H}$ and $\bar{n_{\rm H}}$ are the hydrogen density and its cosmic mean, $\bar{n_{\rm H}}=1.77\times10^{-7}(1+z)^3~{\rm cm^{-3}}$, respectively. Thus, the gas was possibly in a form of a warm, diffuse intergalactic medium as predicted by the numerical simulations (e.g., \cite{Cen99}).

In addition to heating caused by gas compression by gravitational confinement, non-gravitational heating is a plausible mechanism, particularly in smaller systems, such as a group of galaxies.  Therefore, we consider herein the possibility of an AGN feedback by evaluating the excess gas entropy at the cluster center $\Delta K_0$. 
We first estimated rest-frame K-band luminosity ($L_{\rm K}$) of the brightest cluster galaxy near the cluster center based on SED template fitting provided by the COSMOS2015 catalog \citep{Laigle16} for objects in the COSMOS field and the MIZUKI photo-z catalog \citep{Tanaka15,Tanaka18} for objects in the DEEP2-F3 field. We then converted $L_{\rm K}$ to $\Delta K_0$ based on the $L_K-\Delta K_0$ relation \citep{Wang10}. We found $\Delta K_0$ to be on the order of $10^{-2}~{\rm keV\,cm^2}$ in most of the blue clusters. This was negligible in comparison with the baseline model of the gravitational heating \citep{Voit05}, suggesting that the AGN feedback of the central galaxy is not important in the present sample.

The preheating model predicts that the energy input through galactic finds and outflows powered by supernovae should also cause a diffuse gas distribution, resulting in a gas density being too low to be detected in X-rays \citep{Ponman99}. The observed low gas density and as the high emitter fraction are not in conflict with this view, however, the limited photon statistics does not allow us to put any quantitative constraint on the galaxy-ICM connection.

\subsection{Evolutionary stage of the blue clusters}
Based on the above discussion, we suggest that the blue clusters are vitalizing, young systems, and that gas is yet to be compressed and heated to emit appreciable X-rays. Follow-up spectroscopic observations of member galaxies will help directly probe the cluster dynamical status.

The WL analysis and the X-ray-optical comparison indicated that the blue galaxies are small clusters or groups that roughly follow the luminosity-mass relation of the optically-selected clusters. Furthermore, the amount of gas mass can be interpreted as typical of a $\sim 10^{13-14}~{\rm M_{\odot}}$ cluster; thus, they are not gas-poor but possibly at the early stage of cluster formation. We must establish the mass-observable relation based on the follow-up observations to derive a more conclusive result. We will extend the study to a larger sample to more accurately characterize the mass structure of the blue clusters and improve our understanding of the galaxy-ICM co-evolution.


\begin{ack}
The Hyper Suprime-Cam (HSC) collaboration includes the astronomical communities of Japan and Taiwan, and Princeton University.  The HSC instrumentation and software were developed by the National Astronomical Observatory of Japan (NAOJ), the Kavli Institute for the Physics and Mathematics of the Universe (Kavli IPMU), the University of Tokyo, the High Energy Accelerator Research Organization (KEK), the Academia Sinica Institute for Astronomy and Astrophysics in Taiwan (ASIAA), and Princeton University.  Funding was contributed by the FIRST program from the Japanese Cabinet Office, the Ministry of Education, Culture, Sports, Science and Technology (MEXT), the Japan Society for the Promotion of Science (JSPS), Japan Science and Technology Agency  (JST), the Toray Science  Foundation, NAOJ, Kavli IPMU, KEK, ASIAA, and Princeton University.

 This paper makes use of software developed for the Large Synoptic Survey Telescope. We thank the LSST Project for making their code available as free software at  http://dm.lsst.org

 This paper is based [in part] on data collected at the Subaru Telescope and retrieved from the HSC data archive system, which is operated by Subaru Telescope and Astronomy Data Center (ADC) at NAOJ. Data analysis was in part carried out with the cooperation of Center for Computational Astrophysics (CfCA), NAOJ.

 The Pan-STARRS1 Surveys (PS1) and the PS1 public science archive have been made possible through contributions by the Institute for Astronomy, the University of Hawaii, the Pan-STARRS Project Office, the Max Planck Society and its participating institutes, the Max Planck Institute for Astronomy, Heidelberg, and the Max Planck Institute for Extraterrestrial Physics, Garching, The Johns Hopkins University, Durham University, the University of Edinburgh, the Queen’s University Belfast, the Harvard-Smithsonian Center for Astrophysics, the Las Cumbres Observatory Global Telescope Network Incorporated, the National Central University of Taiwan, the Space Telescope Science Institute, the National Aeronautics and Space Administration under grant No. NNX08AR22G issued through the Planetary Science Division of the NASA Science Mission Directorate, the National Science Foundation grant No. AST-1238877, the University of Maryland, Eotvos Lorand University (ELTE), the Los Alamos National Laboratory, and the Gordon and Betty Moore Foundation.

This work was supported in part by JSPS KAKENHI grants 20K04027 (NO), 18J01050 and 19K14759 (YT).
\end{ack}

\bibliographystyle{pasj}
\bibliography{ref.bib}

\end{document}